\documentclass[12pt]{article}
\usepackage{amsmath}
\usepackage{graphicx}
\usepackage{comment} 
\usepackage{xcolor}
\usepackage{amssymb}
\usepackage{amsthm}

\newcommand{\blind}{0}

\begin{document}

\def\spacingset#1{\renewcommand{\baselinestretch}%
{#1}\small\normalsize} \spacingset{1}


\if0\blind
{
  \title{\bf Noise-induced network bursts and coherence in a calcium-mediated neural network}
  \author{Na Yu$^a$, Gurpreet Jagdev$^a$, Michelle Morgovsky$^b$\\
    {\small $^a$Department of Mathematics, Ryerson University, Toronto, Canada}\\
    {\small $^b$Department of Chemistry and Biology, Ryerson University, Toronto, Canada}\\
    }
\date{}    
  \maketitle
} \fi

\if1\blind
{
  \bigskip
  \bigskip
  \bigskip
  \begin{center}
    {\LARGE\bf Title}
\end{center}
  \medskip
} \fi

\bigskip
\begin{abstract}
Noise induced population bursting has been widely identified to play important roles in information process. We constructed a mathematical model for a random and sparse neural network where bursting can be induced from the resting state by global stochastic stimulus. Importantly, the noise-induced bursting dynamics of this network is mediated by the calcium conductance. We use two spectral measures to evaluate the network coherence in the context of network bursts, the spike trains of all neurons and the individual bursts of all neurons. Our results show that the coherence of the network is optimized by an optimal level of stochastic stimulus, which is known as coherence resonance (CR). We also demonstrate that interplay of calcium conductance and noise intensity can modify the degree of CR.
\end{abstract}

\noindent%
{\it Keywords:  noise-induced burst firing, population bursts, coherence resonance, neural network}  
\vfill

\newpage
\spacingset{1.8} 

\section{Introduction}
\label{sec:intro}


Bursting is one of the fundamental coding strategies for neuronal information processing and transmission in the brain \cite{LISMAN199738,IZHIKEVICH2003,Zeldenrust2018,Williams2021}. Its temporal pattern is characterized by the repetitive switch between a silent phase (with almost no spike emission) and an active phase (with two or more spikes with high firing rates). Network bursts (or population bursts) refer to synchronous or near-synchronous burst firing across a neural network \cite{Fardet2018}. The generation of burst firing is regulated by low-threshold calcium channels in various neuronal populations \cite{Cain2013,joksimovic2017role}.  Many calcium imaging studies reported that such neuronal populations are relatively random, dispersed networks \cite{Takano2012}. The synaptic connectivity between neurons is anti-correlated with their lateral distance \cite{Reimann2017,hellwig2000quantitative}, thus, bursting networks mediated by calcium channels have a rather low connection probability.

Coherence is one of the most common measures used to quantify the correlation or the synchronicity of oscillatory patterns of neurons across a neural network \cite{bowyer2016coherence}.  The collective activities of neural networks are often influenced by a ubiquitous and often significant component---noise \cite{faisal2008noise}. Generally, noise can be either local (independent and uncorrelated for each neuron in the network), or global (identical across the network) \cite{Lindner1995,collins1995stochastic}. It has long been shown that noise can play a constructive role to improve the performance of a dynamical system through, for example, coherence resonance (CR). CR is a resonant mechanism where an appropriate amount of noise alone (i.e. without external periodic stimulus) drives a quiescent but excitable system to produce the most coherent oscillations \cite{gang1993,pikovsky1997}. In neuronal dynamics, oscillations represent the neuronal spikes or bursts and the system represents a neuron, or a network. CR has been observed in neural networks such as globally coupled networks \cite{andreev2018,yilmaz2016autapse, kim2015coherence,reinker2006,stacey2002,wang2000coherence}, randomly connected neural networks exhibiting single oscillations \cite{YU20181201,pham1998noise}, 
small-world networks \cite{sun2008spatial,zheng2008spatiotemporal}, ring networks \cite{Zheng2019, masoliver2017}, multiplex networks \cite{MASOLIVER2021,Yamakou2019,semenova2018cr}, and the influencer network of phase oscillators \cite{tonjes2021}.  However, the stochastic dynamics of a calcium-mediated random and sparse heterogeneous bursting network have not been extensively investigated and the effects of CR in such a network remains elusive.

In this work, we consider a quiescent but excitable network mediated by the calcium current, where the connections between neurons are random and sparse.  More specifically, a biological process, spike-timing-dependent plasticity (STDP) is used to simulate the dynamic synapse between two communicating neurons, where the spike timing information of pre-synaptic and post-synaptic neurons is used to adjust the synaptic strength over time \cite{Shouval2010}. We then show that noise can induce network bursts and increase coherence in such a network. In particular, our analysis demonstrates that at an optimal intensity of the global noise, the similarity relations between neurons are maximized, indicating the occurrence of CR. Moreover, by altering the calcium conductance we explore the impact of the calcium current on network coherence.

The remainder of this paper is organized as follows: Section 2 contains the description of the network model. Section 3.1 introduces the noise-induced bursting generated by this network. Sections 3.2-3.4 focuses on the CR of the network bursts, all spikes, and all bursts, respectively. Section 3.5 examines the effect of the calcium conductance on the network coherence. Discussion is given in Section 4.

\section{Mathematical Model} 
\label{sec:model}
We model a network of $N=100$ neurons with random synaptic connections, which adapts the form of the reduced Morris-Lecar model with a linear slow subsystem \cite{izhikevich2000neural}. 
\begin{align}
C\frac{dv_i}{dt} &= I_i - I_{Ca,i} - I_{K,i} - I_{L,i} + I_{loc,i} - I_{syn,i} + I_{glo} \\ 
\frac{dw_i}{dt} &= \phi \lambda_w(v_i) (w_{\infty}(v_i)-w_i) \\
\frac{dI_i}{dt} &= \epsilon (v_0-v_i) \\
\frac{dg_i}{dt} &= -\frac{g_i}{\tau_{e}}
\end{align}
where $v_i$ is the membrane potential of the $i$th neuron for $i=1,2,3,\ldots,N$. $I_{Ca,i}=g_{Ca} m_{\infty}(v_i) (v_i-v_{Ca})$, $I_{K,i}=g_k w_i (v_i-v_k)$, and $I_{L,i}=g_l(v_i-v_l)$ are the calcium, potassium, and leakage currents, respectively, with gating functions 
\begin{align*}
    m_\infty(v_i) &= \frac{1}{2}\left(1+\tanh{\frac{v_i-v_1}{v_2}}\right),\\
    w_\infty(v_i) &= \frac{1}{2}\left(1+\tanh{\frac{v_i-v_3}{v_4}}\right),\\
    \lambda_w(v_i) &= \frac{1}{3}\cosh{\frac{v_i-v_3}{2v_4}}.
\end{align*} 
$ w_i$ is a gating variable of $I_{K,i}$ with $\phi$ as the scaling rate of channel opening. $I_i$ is the linear feedback input current with feedback coefficient $\epsilon$. $g_i$ is the time-varying conductance of the synaptic current with time constant $\tau_e$. $g_{Ca}$, $g_k$, and $g_l$ are the maximum conductances of the calcium, potassium, and leakage currents, with corresponding reversal potentials $v_{Ca}$, $v_k$, and $v_l$. $I_{loc,i}=D_1 \xi_i$ and $I_{glo}=D_2 \eta$ represent the local intrinsic noise (unique for each neuron) and global external stochastic stimulus (same for all neurons), respectively, where $\xi_i$ and $\eta$ are independent Gaussian white noise with mean 0 and variance 1, and $D_1$ and $D_2$ are scaling parameters for the local and global noise intensities, respectively. The parameter values of this model are listed in Table 1. \par

This network is randomly connected with a probability of 15\%. That is, the connection probability for each pair of neurons is 15\%, which is a realistic assumption for a calcium-sensitive neural network, based on experimental study in \cite{hellwig2000quantitative}. An example synaptic connectivity map is presented in Fig.~1A. The synaptic current for the $i$th neuron, $I_{syn,i}$, in Equ.~(1), is averaged by the number of incoming connections from pre-synaptic neurons. That is, $I_{syn,i} = \frac{1}{N_{pre}} \underset{j\neq i \quad \; \,}{\sum_{j=1}^{N}} p_{ji} g_i(v_j-v_{e})$ where $N_{pre}$ represents the number of incoming synapses to neuron $i$; $p_{ji}=1$ if the $j$th neuron (pre-synaptic) and $i$th neuron (post-synaptic) are connected and 0 otherwise; and $\sum_{i=1}^N\sum_{j=1}^{N} p_{ji}$ = $15\% \times N \times N=1500$. The synapses are modeled by a phenomenological model with a spike-timing-dependent-plasticity (STDP) mechanism \cite{Fuhrmann2002,stimberg2019modeling}, which is an important feature for synaptic memory formation and removal. Particularly, synapse release is defined by the product of the two variables, $x_s$ and $u_s$, which represent the fractions of neurotransmitter available and docked for release, respectively. Between action potentials, $u_s$ and $x_s$ follow the dynamics
\begin{align*}
    \frac{du_s}{dt} &= -\Omega_f u_s, \\
    \frac{dx_s}{dt} &= \Omega_d(1-x_s).
\end{align*}
Whenever a pre-synaptic action potential arrives at a post-synaptic cell, the excitatory conductance increases according to $g_i \leftarrow g_i + w_e u_s x_s$, where $w_e$ is the synaptic weight. \par
 
To simulate our network model, we use the \textit{Brian2} package in Python using the Euler-Maruyama method. We then export the simulation results (i.e. the spike times, global stochastic stimuli, and membrane potentials) and proceed with our analysis using MATLAB. For example, MATLAB is used to compute the peri-stimulus time histogram (PSTH), power spectrum density (PSD) and signal to noise ratios (SNRs). 50 trials are used to average the PSDs and SNRs in Figs.~2-8. 

\begin{table}
\caption{Parameter values  \label{parametertable}}
\begin{center}
\begin{tabular}{ccc|ccc}
\hline \hline
Parameter & Value & Unit & Parameter & Value & Unit\\\hline
$v_0$ & -20 & mV & $v_l$ & -50 & mV \\
$v_1$ & -1 & mV & $g_k$& 1.2 & mS \\
$v_2$ & 15 & mV & $g_l$& 0.6 & mS \\
$v_3$ & 10  &mV &$\phi$ & 1 & 1/ms\\
$v_4$ & 5 & mV & $\epsilon$ & 0.001 &mS/ms\\
$v_{Ca}$ & 90 & mV &  $C$ & 1&$\mu$F\\
$v_k$ & -100 & mV &$w_e$ & 0.25 & mS\\
$\Omega_d$ & 4& 1/s & $v_e$ & 20 & mV\\
$\Omega_f$ & 4 & 1/s & $g_{Ca}$  & 0.63 $\sim$ 0.646 & mS\\
 $\tau_e$ & 0.55 & ms &   &  & \\
\hline
\end{tabular}
\end{center}
\end{table}

\section{Results}

\label{sec:Results}

\subsection{Noise-induced bursting}

This study examines the stochastic dynamics of an excitable neural network, whose population activity is characterized by bursts when a global stochastic stimulus is applied. Its corresponding deterministic network (where $D_1=D_2=0$) is quiescent for $g_{Ca}<0.648$ and exhibits periodic bursts of three or more spikes when $g_{Ca}\geq 0.648$. The proposed network rests in the excitable regime (i.e. $g_{Ca}<0.648$), where bursting spikes are stimulated by local intrinsic noise, $D_1 \xi_i$, or external stochastic input, $D_2 \eta$. The network has two heterogeneous components: the local noise ($I_{loc,i}$) and synaptic currents ($I_{syn,i}$, due to random connections between neurons). When $g_{Ca}$ is slightly lower than the excitation threshold (0.648), bursting spikes can be evoked by local noise alone, as shown by one voltage segment in Fig.~1B with $g_{Ca} = 0.646$, $D_1 = 0.007$, and $D_2 = 0$. Our simulation shows, by taking $g_{Ca} = 0.646$, $D_1 = 0.007$, and $D_2 = 0$, bursts occur for all 50 trials with a low average occurrence rate of population bursts (0.64 bursts/second). When $g_{Ca}$ is much lower than $0.648$, local noise alone can not evoke bursting behaviour (see the voltage segment with $g_{Ca} = 0.642$, $D_1 = 0.007$, and $D_2 = 0$ in Fig.~1B). 

\begin{figure}
\begin{center}
\includegraphics[width=6.4in]{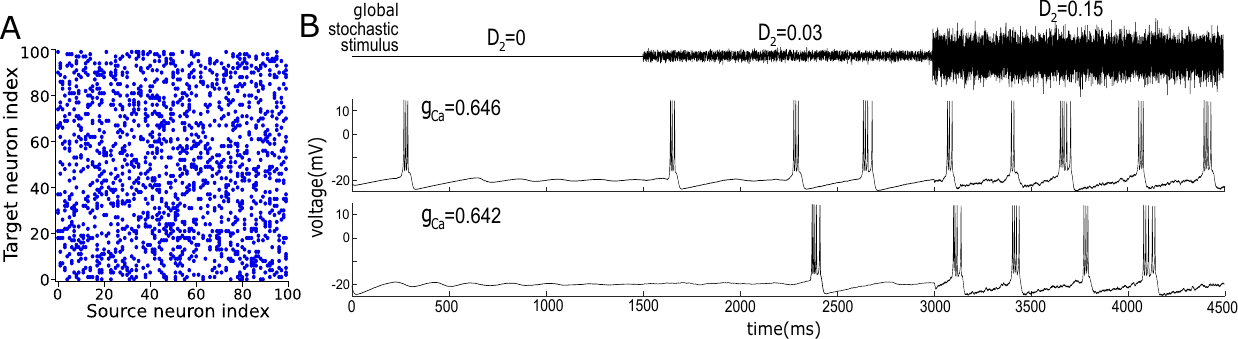}
\end{center}
\caption{(A) An example synaptic connectivity map for a network of $N=100$ neurons with random connection probability of 15\%. The connectivity varies for different trials. (B) Example voltage traces of one neuron in this network in response to a global stochastic stimulus, $D_2 \eta$. This global stochastic stimulus (top row) has three different intensities $D_2=0, 0.03$,  and $0.15$ for every 1500ms segment. Whereas, the intensity of local intrinsic noise, $D_1$, does not change over the three 1500ms segments ($D_1=0.007$). Note, $g_{ca}=0.648$ is the excitation threshold between quiescent and bursting regimes of the deterministic state of this network (where $D_1=D_2=0$). Thus, two calcium conductance's are taken (middle row: $g_{ca}=0.646$ and bottom row: $g_{ca}=0.642$) in order to illustrate the dynamical change of voltage traces.}
\end{figure}

The addition of a global stochastic stimulus, $D_2 \eta$, increases both the burst-generation probability and the occurrence rate of population bursts. As shown by the voltage traces in Figure 1B, when the stochastic stimulus changes from a weak level ($D_2=0.03$) to a relatively higher level ($D_2=0.15$), burst rate increases. Meanwhile, for a larger $D_2$, the number of spikes in a single burst event becomes more random. For example, Fig.~2B (middle row) shows that most of bursts have 3 spikes when $D_2$=0.03, whereas, when $D_2=0.15$ the number of spikes within one burst ranges from 2 to 5. Another observation is that the increment of $D_2$ causes a higher voltage fluctuation on the slow silent phase in between consecutive burst events (i.e. the hyperpolarization stage of action potentials where voltage are around -27 mV to -16 mV), as illustrated in Fig. 1B with $D_2$=0.15. The fluctuations on the slow silent phase can be used to determine the intensity level of global stochastic input, for example, $D_2$=0.03 and 0.15 correspond to weak and intermediate levels, respectively.

\subsection{Coherence Resonance (CR) of network bursts}

 To study network dynamics---and by extension network coherence---we subject our network to weak, intermediate, and strong levels of the global stochastic stimulus, $D_2\eta$. The change in network bursting can be visualized by raster plots and peri-stimulus time histograms (PSTH), as shown in Fig.~2A-2C. A raster plot is a collection of the spike times of individual neurons in a network, where each black dot in the raster plot represents a spike. The PSTH (blue curves in Fig.~2A-2C) summarizes the number of spikes from all neurons across the network at a certain time. Therefore, it records the timing of network bursts, and the height and width of the PSTH peaks indicate the synchrony of individual spikes within a burst. When a rather weak stimulus is applied, the network produces less population bursts. For example, there are 4 population bursts over a 1600 ms period with $D_2$=0.1, as shown in Fig. 2A. The height and the width of some of the PSTH peaks are relatively short and wide, respectively, which reflects a relatively low spike-to-spike synchronization within such burst events (see the 2nd burst in Fig. 2A). When the stimulus is increased to an intermediate level (e.g. $D_2=0.225$ in Fig. 2B), bursting activity becomes more frequent and the PSTH has higher and narrower peaks; spiking events are tightly contained in bursts and the network becomes highly synchronized. $D_2=0.225$ is chosen here because it is the optimal stimulus intensity of CR (as in Fig. 4A). However, when the stimulus is further increased to stronger levels (e.g. $D_2=0.4$ in Fig. 2C), bursting becomes more frequent, but synchronization is destroyed. The sharp peaks of the PSTH become broader, indicating that noise starts to overpower the network dynamics.
 
 \begin{figure}
\begin{center}
\includegraphics[width=5in]{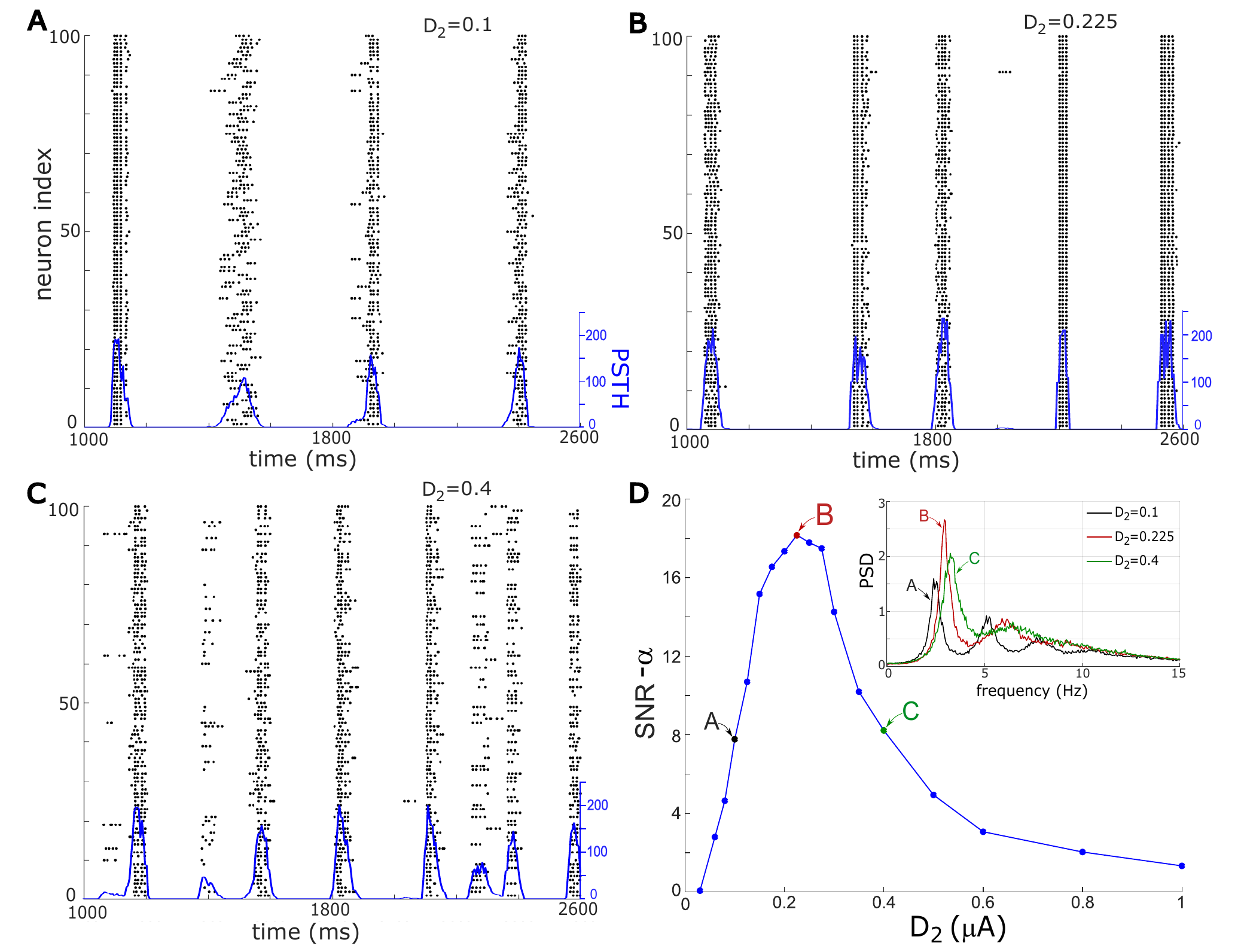} 
\end{center}
\caption{(A)-(C) The spike raster plots (black dots) and the peri-stimulus time histogram (PSTH, blue curves) for $D_2$=0.1, 0.225 and 0.4 (representing weak, intermediate, and strong stimulus levels, respectively) when $g_{Ca}=0.64$. The vertical axis corresponds to the index of a neuron in the network. Each dot indicates that one neuron generates a spike at the time corresponding to the horizontal (time) axis. The bin-width of PSTH is 20 ms. (D) Signal-to-noise ratio (SNR) $\alpha$ v.s. the intensity of stochastic stimulus $D_2$. SNR-$\alpha$ is defined in Equ. (5). Letters A, B and C correspond to $D_2$ values as in panels (A)-(C), which represent three noise intensities of global stimulus: weak; optimal; and strong. Inset of (D): the power spectral density (PSD) of PSTH v.s. frequency for $D_2=0.1$, 0.225 and 0.4. PSD are computed based on the PSTH and averaged over 50 trials. Letters A-C beside the PSD curves indicate their corresponding example raster plots and PSTH in panels (A)-(C). }
\end{figure}

 Such temporal change of network dynamics may also be viewed by the power spectral density (PSD) in the frequency domain. PSTHs are used to calculate the PSD and the average PSD over 50 trials is shown in the inset of Fig. 2D. The black, red, and green PSD curves are labelled by letters A, B, and C, and correspond to Fig. 2A (weak stimulus case), Fig. 2B (intermediate/optimal stimulus case), and Fig. 2C (strong stimulus case), respectively. Three major features of the PSD are often considered. The first is the central frequency (or called resonant frequency), which is the frequency location of the highest PSD point, and it is the reciprocal of the average inter-burst interval (IBI). The central frequency increases with $D_2$, which agrees with the network dynamics in the time domain (Fig. 2A-2C) where the IBI is smaller with the increment of $D_2$. The central frequency is also positively correlated with the burst rate since a shorter IBI implies a higher burst rate. The second and third PSD features are the height and half width of PSD peaks. It is obvious that the optimal stochastic stimulus results in the most pronounced PSD peak (red curve in Fig. 2D inset, with the largest height and the smallest half-width) as opposed to the cases of weak and strong stochastic stimulus (black and green curves in Fig. 2D). This is caused by the higher and narrower PSTH peaks as shown in Fig. 2B.

The network dynamics observed in Fig. 2A-2C and the inset of Fig. 2D indicate that the intensity of the global stochastic stimulus plays an important role in modifying the coherence of our network. To measure coherence more concretely we use the signal-to-noise ratio (SNR) measure \cite{gang1993,pikovsky1997}, 
\begin{equation}
    \alpha= h_p(\Delta \omega/\omega_p)^{-1}, \label{SNRalpha}
\end{equation} 
where $h_p$ and $\omega_p$ denote the height and central frequency of the PSD peak, respectively, and $\Delta\omega$ denotes the width of the PSD peak at half maximal power. For this network, $\omega_p$ has a rather slight change as $D_2$ increases (see Fig. 2D inset), so the ratio between $h_p$ and $\Delta \omega$ dominates $\alpha$. As discussed, the PSD peaks are most pronounced (i.e. large $h_p$ and small $\Delta\omega$) at the intermediate stimulus values as opposed to the weak and strong stimulus values, therefore the SNR-$\alpha$ will peak at intermediate levels of the stochastic stimulus. The SNR-$\alpha$ curve is presented in Fig. 2D. One sees that for weak stimulus the SNR rapidly increases, reaches a maximum at intermediate stimulus values, and then decreases and tends toward zero for strong stimulus values. This a characteristic pattern of CR \cite{gang1993,pikovsky1997} and the peak of the SNR curve corresponds to the maximum degree of network coherence. The intensity of stochastic stimulus which maximizes SNR is called the optimal intensity, and for $g_{ca}=0.64$ in Fig. 2D, the optimal intensity is $D_2=0.225$.

\subsection{The network coherence in terms of all spikes}

PSTH is a collective quantity describing population bursts and does not accurately capture the fast dynamics of intra-burst spikes (i.e. individual spikes within a burst), and as a result, its PSD (and coherence measure SNR-$\alpha$) covers only the low-frequency range of 0-15 Hz. To study the network coherence, we must consider larger bandwidths of frequency to account for both bursts and intra-burst dynamics (i.e. both fast and slow dynamics). Therefore, we will analyze the spike trains of individual neurons in this network. The histogram of inter-spike intervals (ISIs) of spike trains shows a bimodal distribution (two separate and independent peaks): one peak is located at shorter ISIs corresponding to the fast intra-burst dynamics and the other one is at longer ISIs due to the slow dynamics of burst events. As shown in Fig. 3A, with $g_{Ca}=0.64$ and $D_2=0.05$, the majority of intra-burst ISIs are concentrated around 10 ms and inter-burst ISIs are located in the interval of [345, 415] ms. $D_2$=0.05 is chosen because it is the optimal stimulus intensity of CR as shown later in Fig. 4A. This temporal feature of spike trains can also be illustrated by the PSD as shown in Fig. 3B. The PSD peak at low frequency (around 2.5 Hz) corresponds to the longer inter-burst ISIs, and the PSD peak at high-frequency (around 100Hz) results from the shorter intra-burst ISIs, whereas the PSDs of PSTH in the inset of Fig. 2D do not have a peak at a frequency range higher than 15 Hz.  

\begin{figure}
\begin{center}
\includegraphics[width=4in]{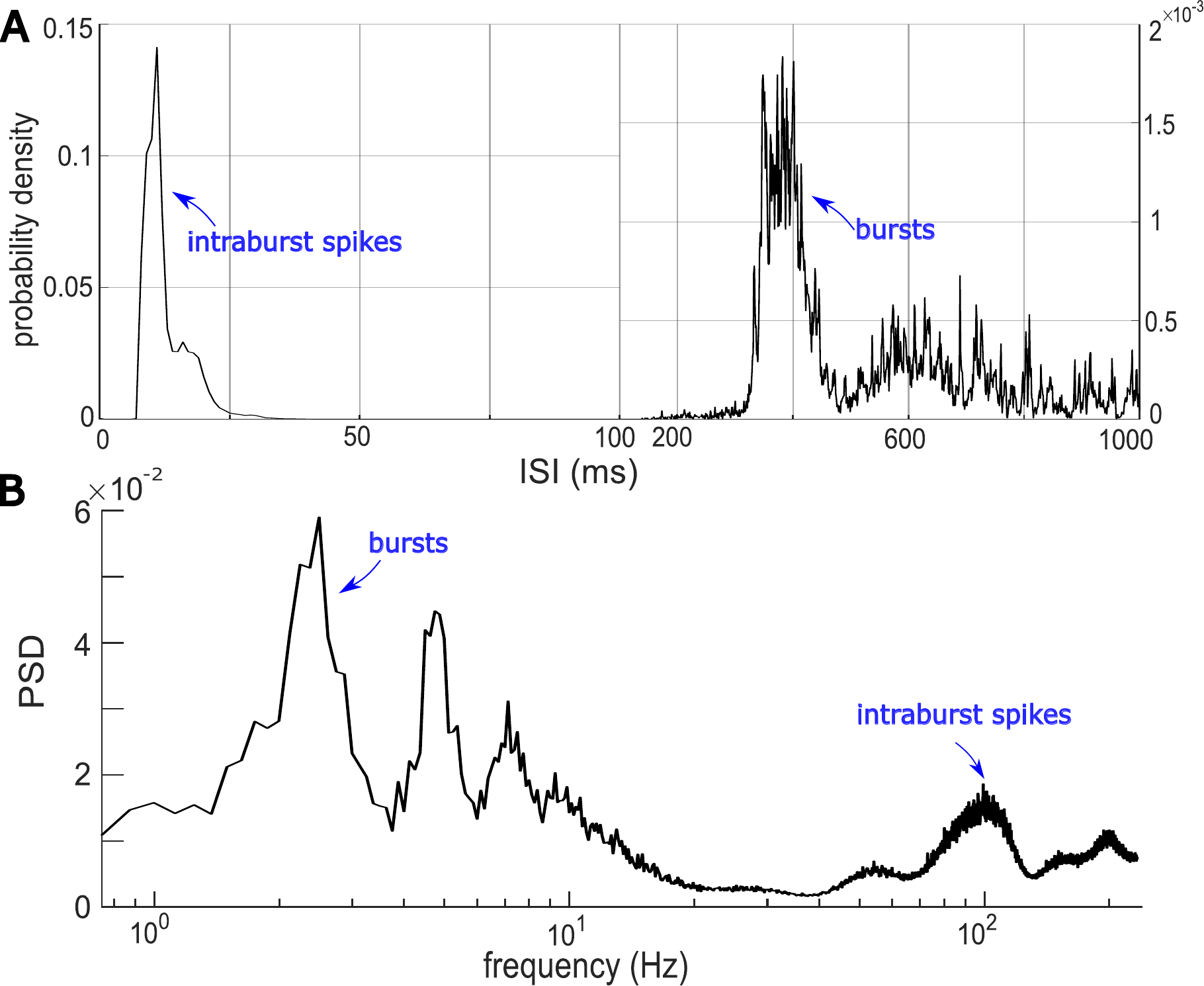}
\end{center}
\caption{(A) The inter-spike interval histogram (ISIH) when $g_{Ca}=0.64$ and $D_2=0.05$. In order to clearly show both intraburst spikes and burst events, different horizontal- and vertical- scales are used on the left and right parts in (A). (B) PSD v.s. frequency when $g_{Ca}=0.64$ and $D_2=0.05$. Note, a logarithmic scale on the horizontal axis is used to show both intraburst spike and burst events. Panel (B) is the representation of panel (A) in the frequency domain.}
\end{figure}

To get a more precise insight of the coherence of all spikes (from both low- and high-frequency ranges) in one network, we use input-output SNR measure \cite{rieke1999spikes}: the ratio between the power of spike trains (output) and the power of the global stochastic stimulus (input). This SNR measure is commonly used to select the best recording location through spike sorting, and also to assess the reliability of neural information transmission \cite{schultz2007signal}. To distinguish from the first SNR measure in Equ. \eqref{SNRalpha}, we denote the second SNR measure as $\beta$. That is,
\begin{equation}
 \beta = \frac{P_{output}}{P_{input}} = \frac{1}{N} \sum_{i=1}^{N}\frac{P_{ST,\,i}}{P_N}, \label{SNRbeta}
\end{equation}
where $P_{ST,\,i}$ is the power of the spike train of the $i$th neuron, and $P_N$ is the power of global stochastic stimulus, $D_2\eta$. $D_2\eta$ is identical for all neurons in the network, and thus \eqref{SNRbeta} can be equivalently written as
\begin{equation*}
    \beta = \frac{\frac{1}{N} \sum_{i=1}^{N}P_{ST,\,i}}{P_{N}}.
\end{equation*}
We then introduce the re-scaled PSD which helps us to understand how $P_{ST,\, i}$ and $P_N$ affect SNR-$\beta$. The re-scaled PSD of spike trains is defined by the averaged PSD over $N$ spike trains in a network divided by $P_N$, i.e. 
\begin{equation}
\tilde{S}(f)=\frac{ \frac{1}{N} \sum_{i=1}^{N} S_i(f)} {P_N}, \label{rPSD}
\end{equation}
where $f$ is the frequency, $S_i(f)$ is the PSD of the spike train generated by the $i$th neuron, and $\tilde{S}(f)$ is the re-scaled PSD. $P_N$ is proportional to $D_2$ because $\eta$ takes the form of white noise, which has a constant PSD. The re-scaled PSDs for three $D_2$ values (0.03, 0.05, and 0.08) are demonstrated in the inset of Fig. 4A. Similar to the PSD in Fig. 3B, the re-scaled PSDs have peaks at both low- and high- frequency ranges. In particular, the re-scaled PSD corresponding to $D_2=0.05$ (red curve in Fig. 4A inset) is higher than the other two PSD curves, consequently, SNR-$\beta$ is expected to be larger at $D_2=0.05$. 

The SNR-$\beta$ as a function of $D_2$ is illustrated in Fig. 4A for $g_{Ca}=0.64$ and it also shows a characteristic pattern of CR, with a maximum at $D_2=0.05$. Comparing two coherence measures (SNR-$\alpha$ in Fig. 2D and SNR-$\beta$ in Fig. 4A), two major differences are observed: (a) the noise intensity range is [0.03, 1] for SNR-$\alpha$, but CR measured by SNR-$\beta$ occurs over the weak intensity range of [0.001, 0.3]; and (b) the optimal intensity is 0.225 in Fig. 2D but it is 0.05 in Fig. 4A. The differences above are due to the different focus of the two SNR functions,  SNR-$\alpha$ in Equ. \eqref{SNRalpha} and SNR-$\beta$ in Equ. \eqref{SNRbeta}. Although they both evaluate CR, SNR-$\alpha$ characterizes the similarity of the frequency content of neuronal oscillations (i.e. bursts across the network here), while SNR-$\beta$ focus on the reliability of the neuronal responses (i.e. all spikes across network here) to the input (stochastic stimulus) over time. 

The example raster plots presented in Fig. 4B-4D demonstrate the spatio-temporal patterns of firing for this neural network with respect to the increment of $D_2$ from 0.03 to 0.08. When subjected to very weak levels stimulus (e.g. $D_2=0.03$ in Fig. 4B) bursting is induced at a very low rate (around 0.6 bursts/second) with a very low synchronization. When stimulus is slightly increased, for example $D_2=0.05$ in Fig. 4C, both population bursts and individual spikes become more frequent. The network becomes synchronized, spiking events are tightly grouped in bursts, and the SNR-$\beta$ reaches its peak as expected based on the analysis of re-scaled PSDs. This also agrees with the sharp peaks in the histogram of the ISIs (Fig. 3A). When the network is subjected to higher---but still relatively weak---levels of stimulus (e.g. $D_2=0.08$ in Fig. 4D) the stochastic stimulus becomes overpowering and some network bursts start to become dis-synchronized. The observations above are in line with the observations for SNR-$\alpha$ in Fig. 2, although the $D_2$ values differ. 

\begin{figure}
\begin{center}
\includegraphics[width=5in]{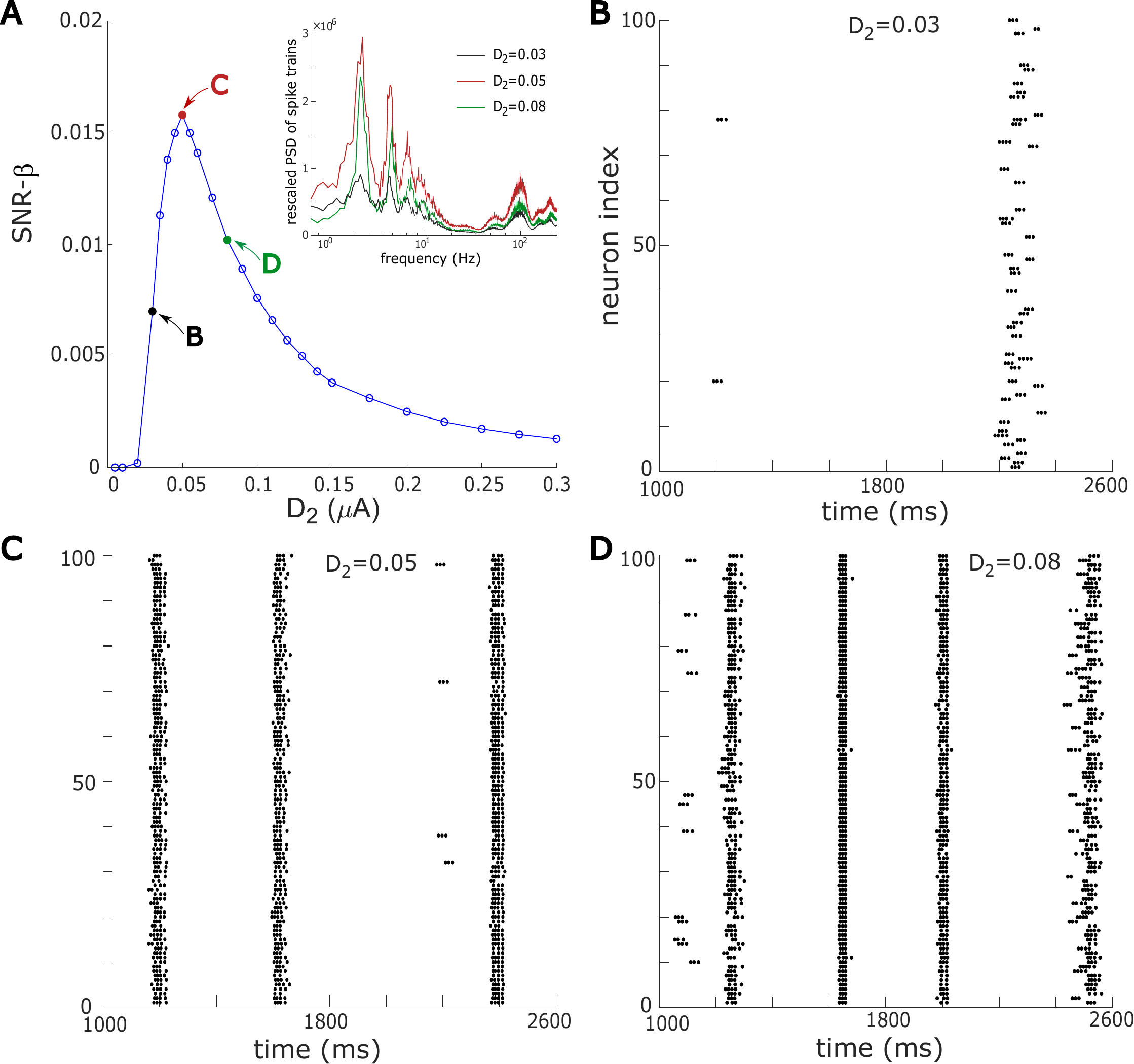}
\end{center}
\caption{(A) SNR-$\beta$ calculated by the power spectra of spike trains v.s. $D_2$ with $g_{Ca}=0.64$. SNR curve reaches a peak (red solid circle) at $D_2=0.05$. Inset of (A): The re-scaled PSDs of spike trains in the frequency domain. The re-scaled PSD is defined in Equ. \eqref{rPSD}. (B)-(C) show three example spike raster plots for the three $D_2$ values as labeled in (A), and other parameter values are as same as in Fig. 2}
\end{figure}

\subsection{The network coherence in terms of individual bursts}

The PSTH estimates the timing of population bursts across a network. In order to accurately record the onset (i.e. the occurrence times) of the bursts produced by individual neurons, we mark a burst event by its initial spike time. Thus a burst train can be formed for each neuron. A burst train is a binary sequence which takes the value 1 at initial spikes in all bursting events and 0 otherwise, as demonstrated in the inset of Fig. 5A. The occurrence of bursts is identified using the dynamic burst threshold method in \cite{selinger2007methods}. 
 

As shown in Fig. 5A, the average burst rate produced by each neuron in a network increases with respect to $D_2$, which is inline with the voltage time series of a single neuron (Fig. 1B). To quantify the coherence of the burst trains, we use the input-output SNR measure similar to Equ. \eqref{SNRbeta}, that is,
\begin{equation}
 \beta = \frac{1}{N} \sum_{i=1}^{N}\frac{P_{BT,\,i}}{P_N}, \label{SNRBT}
\end{equation} 
where $P_{BT,\,i}$ denotes the power of the burst train generated by the $i$th neuron. Fig. 5B illustrates the change of SNR on different $D_2$ for $g_{Ca}=0.64$ (i.e. the same parameter values as Fig. 4A). The SNR-$\beta$ curve in Fig. 5B shows the characteristics of CR: it rapidly increases for very weak levels of the global stochastic stimulus ($D_2<0.05$ in Fig. 5B), reaches a peak value at $D_2=0.05$ in Fig. 5B, and then decreases for larger stimulus intensities ($D_2>0.05$ in Fig. 5B). The SNR-$\beta$ curves calculated from burst trains and spikes trains share the same optimal noise intensity, but it is different from the optimal intensity presented in SNR-$\alpha$ calculated from PSTH (Fig. 2D) because a higher $D_2$ leads to a lower re-scaled PSD curve (Fig. 4A inset) and consequently SNR-$\beta$ is continuously decreasing for $D_2>0.05$. 

\begin{figure}
\begin{center}
\includegraphics[width=5in]{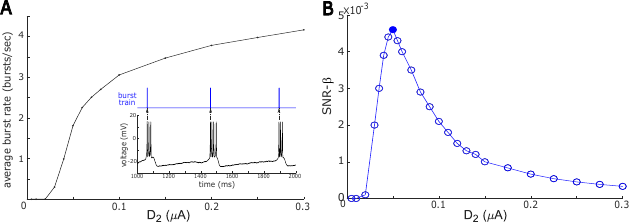}
\end{center}
\caption{(A) Average burst rate produced by each neuron in a network v.s. $D_2$ for $g_{Ca}=0.64$. Inset of (A): an example binary burst train, where the burst onset is defined by the firing time of the first spike of a burst event in the corresponding voltage trace. (B) SNR-$\beta$ curve calculated by the power spectra of the burst trains for $g_{Ca}=0.64$, and other parameter values are same as in Fig. 2. The largest SNR is marked by a solid dot and corresponds to $D_2=0.05$.}
\end{figure}

\subsection{The effect of calcium conductance on CR}

As demonstrated in subsection 3.1, neural dynamics change with the calcium conductance, $g_{Ca}$. In the preceding subsections we have considered $g_{Ca}=0.64$. To study the effects of $g_{Ca}$ on the network coherence we take various $g_{Ca}$ values in the excitable regime (i.e. $g_{Ca}$<0.648) and use SNR-$\beta$ measure calculated from both spike trains (Equ. \eqref{SNRbeta}) and burst trains (Equ. \eqref{SNRBT}). A series of SNR-$\beta$ optimization curves, corresponding to four parameter values of $g_{Ca}$ (0.638, 0.64, 0.642, and 0.645), are plotted for both the spike trains (Fig. 6A) and burst trains (Fig. 6B). All of the SNR curves display the characteristic pattern of CR, that is, at first they are increasing, reach a peak value, and then decrease toward zero. The SNR curve corresponding to a larger $g_{Ca}$ is above the SNR curve corresponding to a smaller $g_{Ca}$ for $D_2\in$ [0.005, 0.15], indicating that the coherence degree is enhanced by increased $g_{Ca}$ over the weak noise intensity range. Moreover, the maximal degree of coherence (i.e. the height of the SNR peaks) and the corresponding optimal intensities ($D_2$) vary across different $g_{Ca}$ values.

\begin{figure}
\begin{center}
\includegraphics[width=5in]{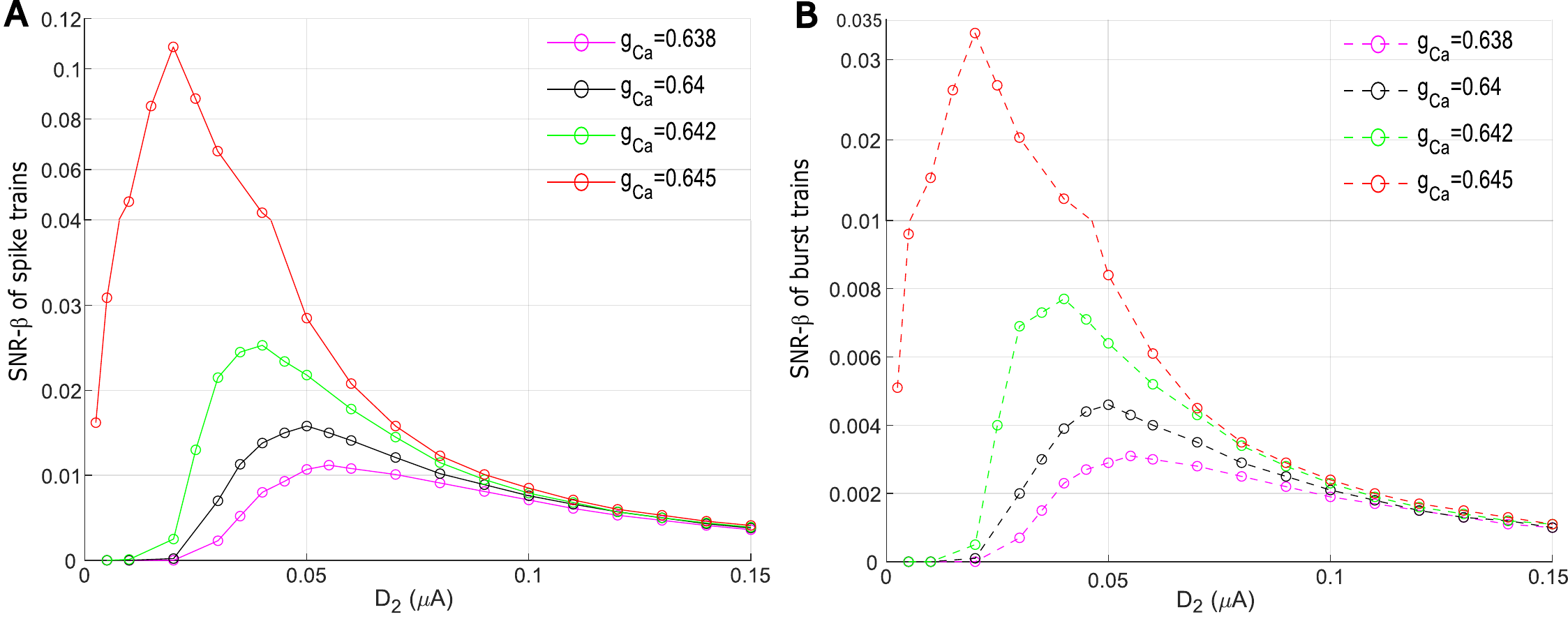}
\end{center}
\caption{(A) SNR-$\beta$ calculated by the power spectra of the spike trains v.s. the intensity of stochastic stimulus, $D_2$. (B) SNR-$\beta$ calculated from the burst trains v.s. $D_2$. For both panels, $g_{Ca}=$ 0.638, 0.64, 0.642 and 0.645, and they are in the excitable regime. In order to clearly show all SNR curves the y-axis has different scales on the top and bottom in both panels.}
\end{figure}

To capture $g_{Ca}$-dependent change in the height of SNRs, we computed the maximum SNR values for seven $g_{Ca}$ values in the excitable regime, ranging from 0.63 to 0.645, and plotted them in Fig. 7A for spike trains and Fig. 7B for burst trains. Four $g_{Ca}$ values in Fig. 6A are part of these seven values. As would be expected, when $g_{Ca}$ is increased (i.e. closer to the excitation threshold), the peak value of SNR increases nearly exponentially. It implies that a higher network coherence is expected at its optimal intensity as the excitable system is in a closer proximity to the excitation threshold.

We also computed the optimal intensities of global stochastic stimulus, $D_2$, for these seven $g_{Ca}$ values. We found that SNRs calculated from the power spectra of spike trains and burst trains share the same optimal intensity  for one $g_{Ca}$; they are plotted as a function of $g_{Ca}$ in Fig. 7C. With increased $g_{Ca}$, the optimal intensity of the SNR decreases. This suggests that when the excitable system is closer to the excitation threshold, a smaller noise intensity is able to drive the network to achieve its best possible coherence.

\begin{figure}
\begin{center}
\includegraphics[width=5in]{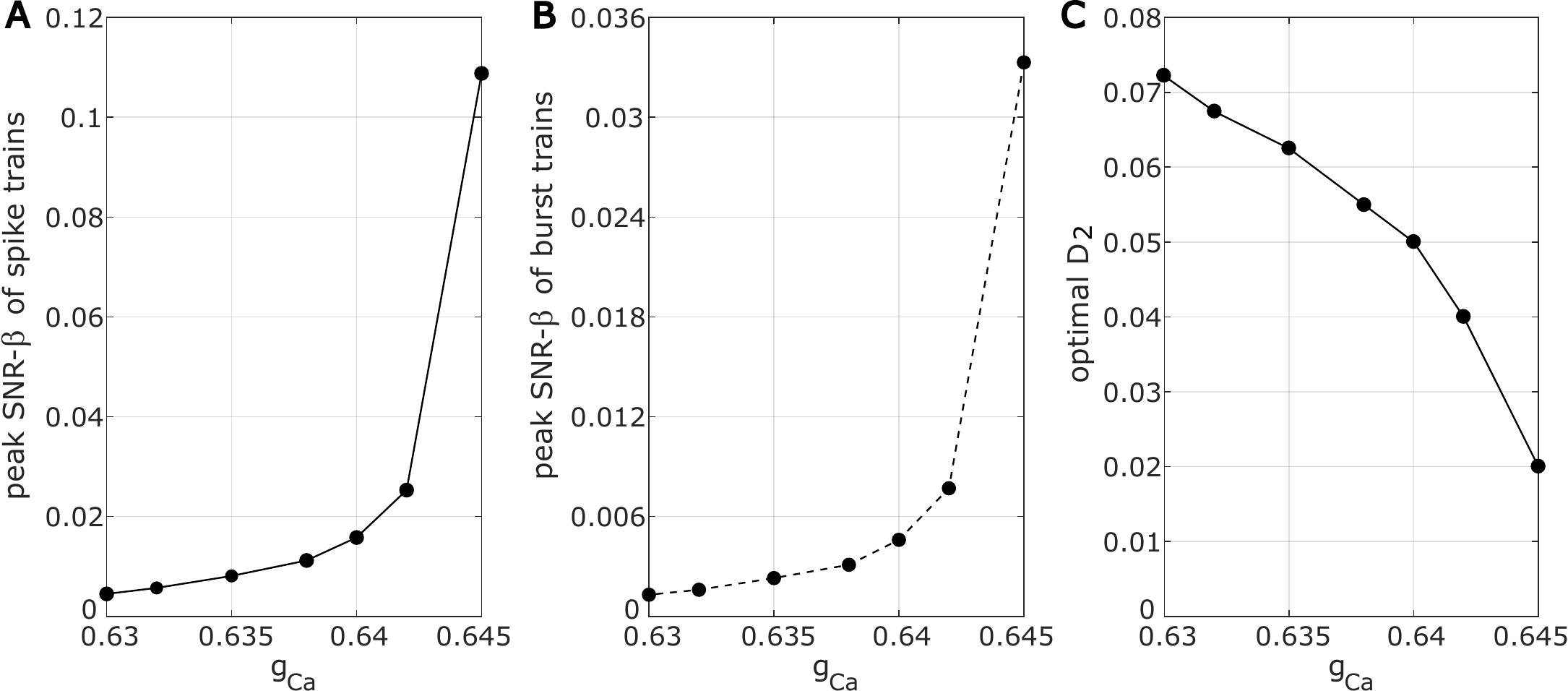} 
\end{center}
\caption{(A) The maximum degree of SNR-$\beta$ calculated by the power spectra of the spike trains v.s. $g_{Ca}$. (B) The maximum degree of SNR-$\beta$ from the burst trains v.s. $g_{Ca}$. (C) The optimal intensities of global stochastic stimulus ($D_2$) v.s. $g_{Ca}$. For all three panels, $g_{Ca}=$ 0.63, 0.632, 0.635, 0.638, 0.64, 0.642 and 0.645; all of which are in the excitable regime.
}
\end{figure}

As CR degree is sensitive to a slight change of $g_{Ca}$, n order to further study the effects of $g_{ca}$ on network coherence, we add a third heterogeneous component to this network. We let $g_{Ca}$ be a uniform random variable ranging from 0.63 to 0.645 (i.e. $g_{ca} \sim U(0.63,0.645)$), so that each neuron in our network may have a different $g_{Ca}$ value. 
SNR-$\beta$ measure is used to evaluate the network coherence, and we calculate the SNR-$\beta$ using the power spectra of both spike and burst trains (see Eqs. \eqref{SNRbeta} and \eqref{SNRBT}). In Fig. 8, one sees that the SNR---for both spike and burst trains---sharply increases from $D_2=0.005$ to $D_2=0.03$, reaches a peak at approximately $D_2=0.03$ and then for larger $D_2$ tends towards 0. In other words, the network displays a resonant behaviour and the optimal stimulus intensity (e.g. $D_2=0.03$ here) would induce the best coherence (with maximal SNR $\approx$ 0.028 for spike trains and 0.008 for burst trains). Compared to the results in Figs. 6-7, this $g_{Ca}$-varied network has similar peak SNR values and optimal intensities to the $g_{Ca}$-fixed network with $g_{Ca}=0.0642$ where SNR $\approx$ 0.0253 for spike trains and 0.0077 for burst and $D_2=0.04$, although the mean $g_{Ca}$ value here is 0.6375. This is because the CR degree increases nearly exponentially with increased $g_{Ca}$ as shown in Fig. 7A-7B.

\begin{figure}
\begin{center}
\includegraphics[width=3in]{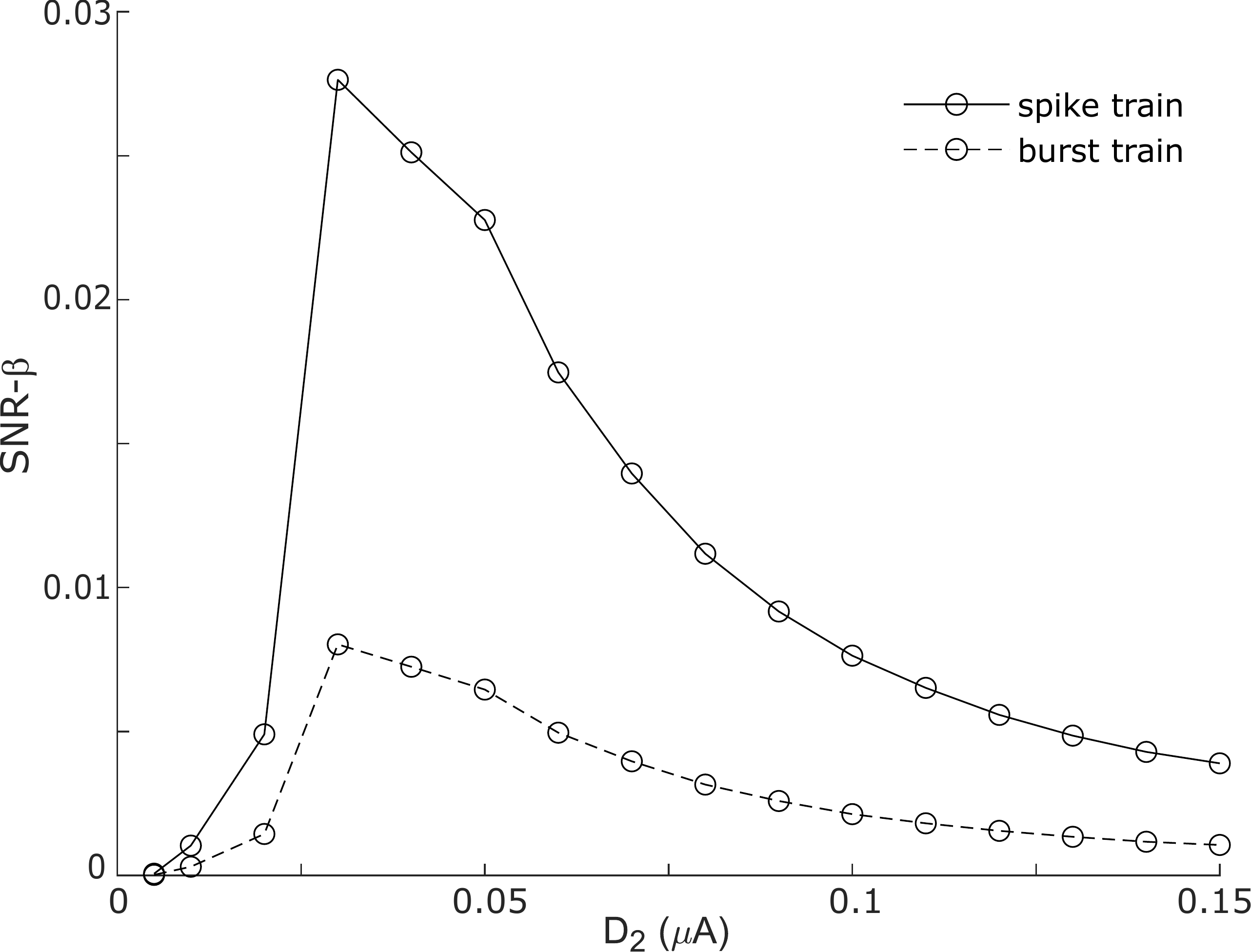}
\end{center}
\caption{SNR-$\beta$ calculated by the power spectra of spike trains (solid line) and burst trains (dashed line) of a $g_{Ca}$-varied network of 100 neurons. A set of 100 uniformly distributed random numbers ranging from 0.63 to 0.645 were used to model different $g_{Ca}$ values in this network. The largest SNR value for both the spike- and burst- trains occurs at $D_2=0.03$.}
\end{figure}
 
\section{Discussion}

Many numerical studies on the stochastic dynamics of neural networks employing homogeneous networks \cite{reinker2006,wang2000coherence,Soula2007}, globally connected networks \cite{andreev2018, yilmaz2016autapse, kim2015coherence,stacey2002,wang2000coherence}. We consider a heterogeneous network mediated by calcium channels, also, it has random and sparse synaptic connections and produces bursts when stimulated by external input. This study provides different views to evaluate the coherence for such a network. In terms of the regularity of the network output, the coherence information extracted from the population bursts indicated that the resonant coherence occurs at the intermediate noise intensity, which agrees with previous studies of globally connected bursting networks (e.g. \cite{kim2015coherence,reinker2006}). This coherence information is helpful for understanding the influences of global noise on the collective dynamics (e.g. population bursts) and the potential physiological functions of a neural network. If network performance is evaluated based on the efficiency of the network response to the stochastic input, a weak stochastic input can stimulate both the coherence of all spikes across the network and the correlation of neuronal bursts to reach the optimal level. This implies that the improvement of neuronal communication within a network can be achieved using weak noise.

The impact of the noise intensity, coupling strength, or the network topology on the CR is often discussed (e.g. \cite{masoliver2017,kim2015}). However, we focus on the effects of the interplay between calcium conductance and noise intensity on CR. Calcium current has been found to regulate the excitation and resonance of individual neurons \cite{Hutcheon1994}. Our work demonstrated that calcium current can also enhance the network response to the stochastic stimuli: when calcium conductance is closer to its excitation threshold, a smaller intensity of stochastic stimulus is needed to induce the best coherence where a higher CR degree is achieved. This gives us hope for experimental discovery of this noise-induced resonance effect by analyzing calcium-related brain response near the excitation threshold.

\bibliographystyle{elsarticle-num}
\bibliography{ref1}

\begin{thebibliography}{10}
\expandafter\ifx\csname url\endcsname\relax
  \def\url#1{\texttt{#1}}\fi
\expandafter\ifx\csname urlprefix\endcsname\relax\def\urlprefix{URL }\fi
\expandafter\ifx\csname href\endcsname\relax
  \def\href#1#2{#2} \def\path#1{#1}\fi

\bibitem{LISMAN199738}
J.~E. Lisman, Bursts as a unit of neural information: making unreliable
  synapses reliable, Trends in Neurosciences 20~(1) (1997) 38--43.

\bibitem{IZHIKEVICH2003}
E.~M. Izhikevich, N.~S. Desai, E.~C. Walcott, F.~C. Hoppensteadt, Bursts as a
  unit of neural information: selective communication via resonance, Trends in
  Neurosciences 26~(3) (2003) 161--167.

\bibitem{Zeldenrust2018}
F.~Zeldenrust, W.~J. Wadman, B.~Englitz, Neural coding with bursts—current
  state and future perspectives, Frontiers in Computational Neuroscience 12
  (2018) 48.

\bibitem{Williams2021}
E.~Williams, A.~Payeur, A.~Gidon, R.~Naud, Neural burst codes disguised as rate
  codes, Scientific Reports 11 (2021) 15910.

\bibitem{Fardet2018}
T.~Fardet, M.~Ballandras, S.~Bottani, S.~Métens, P.~Monceau, Understanding the
  generation of network bursts by adaptive oscillatory neurons, Frontiers in
  Neuroscience 12 (2018) 41.

\bibitem{Cain2013}
S.~M. Cain, T.~P. Snutch, T-type calcium channels in burst-firing, network
  synchrony, and epilepsy, Biochimica et Biophysica Acta (BBA)-Biomembranes
  1828~(7) (2013) 1572--1578.

\bibitem{joksimovic2017role}
S.~M. Joksimovic, P.~Eggan, Y.~Izumi, S.~L. Joksimovic, V.~Tesic, R.~M. Dietz,
  J.~E. Orfila, M.~R. DiGruccio, P.~S. Herson, V.~Jevtovic-Todorovic, et~al.,
  The role of t-type calcium channels in the subiculum: to burst or not to
  burst?, The Journal of physiology 595~(19) (2017) 6327--6348.

\bibitem{Takano2012}
H.~Takano, M.~McCartney, P.~I. Ortinski, C.~Yue, M.~E. Putt, D.~A. Coulter,
  Deterministic and stochastic neuronal contributions to distinct synchronous
  ca3 network bursts, Journal of Neuroscience 32~(14) (2012) 4743--4754.

\bibitem{Reimann2017}
M.~W. Reimann, A.-L. Horlemann, S.~Ramaswamy, E.~B. Muller, H.~Markram,
  {Morphological Diversity Strongly Constrains Synaptic Connectivity and
  Plasticity}, Cerebral Cortex 27~(9) (2017) 4570--4585.

\bibitem{hellwig2000quantitative}
B.~Hellwig, A quantitative analysis of the local connectivity between pyramidal
  neurons in layers 2/3 of the rat visual cortex, Biological cybernetics 82~(2)
  (2000) 111--121.

\bibitem{bowyer2016coherence}
S.~M. Bowyer, Coherence a measure of the brain networks: past and present,
  Neuropsychiatric Electrophysiology 2~(1) (2016) 1--12.

\bibitem{faisal2008noise}
A.~A. Faisal, L.~P. Selen, D.~M. Wolpert, Noise in the nervous system, Nature
  reviews neuroscience 9~(4) (2008) 292--303.

\bibitem{Lindner1995}
J.~F. Lindner, B.~K. Meadows, W.~L. Ditto, M.~E. Inchiosa, A.~R. Bulsara, Array
  enhanced stochastic resonance and spatiotemporal synchronization, Phys. Rev.
  Lett. 75 (1995) 3--6.

\bibitem{collins1995stochastic}
J.~Collins, C.~C. Chow, T.~T. Imhoff, Stochastic resonance without tuning,
  Nature 376~(6537) (1995) 236--238.

\bibitem{gang1993}
H.~Gang, T.~Ditzinger, C.-Z. Ning, H.~Haken, Stochastic resonance without
  external periodic force, Physical Review Letters 71~(6) (1993) 807.

\bibitem{pikovsky1997}
A.~S. Pikovsky, J.~Kurths, Coherence resonance in a noise-driven excitable
  system, Physical Review Letters 78~(5) (1997) 775.

\bibitem{andreev2018}
A.~V. Andreev, V.~V. Makarov, A.~E. Runnova, A.~N. Pisarchik, A.~E. Hramov,
  Coherence resonance in stimulated neuronal network, Chaos, Solitons \&
  Fractals 106 (2018) 80--85.

\bibitem{yilmaz2016autapse}
E.~Yilmaz, M.~Ozer, V.~Baysal, M.~Perc, Autapse-induced multiple coherence
  resonance in single neurons and neuronal networks, Scientific Reports 6~(1)
  (2016) 1--14.

\bibitem{kim2015coherence}
J.~H. Kim, H.~J. Lee, C.~H. Min, K.~J. Lee, Coherence resonance in bursting
  neural networks, Physical Review E 92~(4) (2015) 042701.

\bibitem{reinker2006}
S.~Reinker, Y.-X. Li, R.~Kuske, Noise-induced coherence and network
  oscillations in a reduced bursting model, Bulletin of mathematical biology
  68~(6) (2006) 1401--1427.

\bibitem{stacey2002}
W.~C. Stacey, D.~M. Durand, Noise and coupling affect signal detection and
  bursting in a simulated physiological neural network, Journal of
  neurophysiology (2002).

\bibitem{wang2000coherence}
Y.~Wang, D.~T. Chik, Z.~Wang, Coherence resonance and noise-induced
  synchronization in globally coupled hodgkin-huxley neurons, Physical Review E
  61~(1) (2000) 740.

\bibitem{YU20181201}
H.~Yu, L.~Zhang, X.~Guo, J.~Wang, Y.~Cao, J.~Liu, Effect of inhibitory firing
  pattern on coherence resonance in random neural networks, Physica A:
  Statistical Mechanics and its Applications 490 (2018) 1201--1210.

\bibitem{pham1998noise}
J.~Pham, K.~Pakdaman, J.-F. Vibert, Noise-induced coherent oscillations in
  randomly connected neural networks, Physical Review E 58~(3) (1998) 3610.

\bibitem{sun2008spatial}
X.~Sun, M.~Perc, Q.~Lu, J.~Kurths, Spatial coherence resonance on diffusive and
  small-world networks of hodgkin--huxley neurons, Chaos: An Interdisciplinary
  Journal of Nonlinear Science 18~(2) (2008) 023102.

\bibitem{zheng2008spatiotemporal}
Y.~H. Zheng, Q.~S. Lu, Spatiotemporal patterns and chaotic burst
  synchronization in a small-world neuronal network, Physica A: Statistical
  Mechanics and its Applications 387~(14) (2008) 3719--3728.

\bibitem{Zheng2019}
C.~Zheng, A.~Pikovsky, Stochastic bursting in unidirectionally delay-coupled
  noisy excitable systems, Chaos: An Interdisciplinary Journal of Nonlinear
  Science 29~(4) (2019) 041103.

\bibitem{masoliver2017}
M.~Masoliver, N.~Malik, E.~Sch{\"o}ll, A.~Zakharova, Coherence resonance in a
  network of fitzhugh-nagumo systems: Interplay of noise, time-delay, and
  topology, Chaos: An Interdisciplinary Journal of Nonlinear Science 27~(10)
  (2017) 101102.

\bibitem{MASOLIVER2021}
M.~Masoliver, C.~Masoller, A.~Zakharova, Control of coherence resonance in
  multiplex neural networks, Chaos, Solitons \& Fractals 145 (2021) 110666.

\bibitem{Yamakou2019}
M.~E. Yamakou, J.~Jost, Control of coherence resonance by self-induced
  stochastic resonance in a multiplex neural network, Phys. Rev. E 100 (2019)
  022313.

\bibitem{semenova2018cr}
N.~Semenova, A.~Zakharova, Weak multiplexing induces coherence resonance,
  Chaos: An Interdisciplinary Journal of Nonlinear Science 28~(5) (2018)
  051104.

\bibitem{tonjes2021}
R.~T{\"o}njes, C.~E. Fiore, T.~Pereira, Coherence resonance in influencer
  networks, Nature Communications 12~(1) (2021) 1--8.

\bibitem{Shouval2010}
H.~Shouval, S.~Wang, G.~Wittenberg, Spike timing dependent plasticity: A
  consequence of more fundamental learning rules, Frontiers in Computational
  Neuroscience 4 (2010) 19.

\bibitem{izhikevich2000neural}
E.~M. Izhikevich, Neural excitability, spiking and bursting, International
  journal of bifurcation and chaos 10~(06) (2000) 1171--1266.

\bibitem{Fuhrmann2002}
H.~M. G.~Fuhrmann, I.~Segev, M.~Tsodyks, Coding of temporal information by
  activity-dependent synapses, J Neurophysiol. 87~(1) (2002) 140--148.

\bibitem{stimberg2019modeling}
M.~Stimberg, D.~F. Goodman, R.~Brette, M.~De~Pitt{\`a}, Modeling neuron--glia
  interactions with the brian 2 simulator, in: Computational glioscience,
  Springer, 2019, pp. 471--505.

\bibitem{rieke1999spikes}
F.~Rieke, D.~Warland, Spikes: exploring the neural code, Cambridge, Mass. : MIT
  Press, 1999.

\bibitem{schultz2007signal}
S.~R. Schultz, Signal-to-noise ratio in neuroscience, Scholarpedia 2~(6) (2007)
  2046.

\bibitem{selinger2007methods}
J.~V. Selinger, N.~V. Kulagina, T.~J. O'Shaughnessy, W.~Ma, J.~J. Pancrazio,
  Methods for characterizing interspike intervals and identifying bursts in
  neuronal activity, Journal of neuroscience methods 162~(1-2) (2007) 64--71.

\bibitem{Soula2007}
H.~Soula, C.~C. Chow, {Stochastic Dynamics of a Finite-Size Spiking Neural
  Network}, Neural Computation 19~(12) (2007) 3262--3292.

\bibitem{kim2015}
S.-Y. Kim, W.~Lim, Noise-induced burst and spike synchronizations in an
  inhibitory small-world network of subthreshold bursting neurons, Cognitive
  neurodynamics 9~(2) (2015) 179--200.

\bibitem{Hutcheon1994}
Y.~Y. B.~Hutcheon, R.M.~Miura, E.~Puil, Low-threshold calcium current and
  resonance in thalamic neurons: a model of frequency preference, J
  Neurophysiol 71~(2) (1994) 583--94.

\end{thebibliography}

\end{document}